\documentclass{llncs}
\usepackage{scrextend}
\usepackage{booktabs}
\usepackage{multirow}
\usepackage{float}
\usepackage{tabularx}
\usepackage[table]{xcolor}
\usepackage{graphicx}
\usepackage{array}
\usepackage[flushleft]{threeparttable}
\begin{document}

\title{Do users talk about the software in my product? Analyzing user reviews on IoT products}
%
%
\author{Kamonphop Srisopha \and Pooyan Behnamghader \and Barry Boehm}
%
%
%
\institute{University of Southern California, Los Angeles CA 90089, USA\\
\email{\{srisopha,pbehnamg,boehm\}@usc.edu}}

\maketitle              

\begin{abstract}
Consumer product reviews are an invaluable source of data because they contain a wide range of information that could help requirement engineers to meet user needs. Recent studies have shown that tweets about software applications and reviews on App Stores contain useful information, which enable a more responsive software requirements elicitation. However, all of these studies' subjects are merely software applications. Information on system software, such as embedded software, operating systems, and firmware, are overlooked, unless reviews of a product using them are investigated. Challenges in investigating these reviews could come from the fact that there is a huge volume of data available, as well as the fact that reviews of such products are diverse in nature, meaning that they may contain information mostly on hardware components or broadly on the product as a whole. Motivated by these observations, we conduct an exploratory study using a dataset of 7198 review sentences from 6 Internet of Things (IoT) products. Our qualitative analysis demonstrates that a sufficient quantity of software related information exists in these reviews. In addition, we investigate the performance of two supervised machine learning techniques (Support Vector Machines and Convolutional Neural Networks) for classification of information contained in the reviews. Our results suggest that, with a certain setup, these two techniques can be used to classify the information automatically with high precision and recall.
\end{abstract}
\section{Introduction}

Online retail stores (\textit{e.g.,} Amazon or eBay) provide a convenient platform for producers to compete and sell their products. 
They also provide consumers with the ability to compare a variety of options and purchase the one that best fits their budgets and needs. Consumers can share their experience of using a product through writing a review after purchase. 
This helps other consumers to have a better understanding of the quality of the products. 

Not only do reviews help consumers make more informed decisions, but it also helps manufacturers get feedback on their products so that they can improve them over time. 
In fact, reviews contain a wealth of information that can be used to create new requirements. The idea of mining user reviews to improve the quality of the product is a form of ``Passive Crowdsourcing." According to \cite{howe2006rise}, Crowdsourcing ``represents the act of a company or institution taking a function once performed by employees and outsourcing it to an undefined (and generally large) network of people in the form of an open call."
Passive Crowdsourcing simply means harnessing the power of the crowd without their knowledge. 

Mining app store reviews to improve the quality of software has already gained the attention of researchers \cite{genc2017systematic}. In comparison with user reviews of a software application, user reviews of an IoT product are not all about software. They can be about different product elements, such as software, hardware, customer service, etc. In particular, they contain information such as subjective product evaluations, personal experiences, problems encountered, product descriptions, and users' visions on how the product should be. As a result, reviews of IoT products are more diverse. If such information in this huge volume of data is detected and classified correctly and efficiently, different stakeholders, such as software engineers, hardware engineers, etc. can better meet user expectations and needs, thus accelerating product or software evolution processes.

Motivated by these observations, we conducted an exploratory study by analyzing 1491 verified purchase reviews (7198 review sentences) of 6 IoT products to investigate how IoT product reviews can be categorized, whether enough software related information exists, and how much of this information is useful for software developers and requirement engineers. We also studied the extent to which two supervised machine learning techniques (SVM and CNN) can be used to classify information in the dataset automatically, and identified configurations that can effectively improve the classification performance. We believe that the work presented in this paper contributed to the growing knowledge in this area.

The rest of the paper is organized as follows. We formally define our research problems and describe data collection in Section 2. In Section 3, we describe the procedure to answer each research question and report the results. Thereafter in Section 4, we present factors that may affect the validity of our study. We discuss the related literature in Section 5. Finally, in Section 6, we conclude the paper and outline plans for future research.
\section{Research Design}
\subsection{Research Questions}
To guide our research, we formulate the following research questions:
\begin{itemize}
  \item[] \textbf{RQ1: How can IoT product reviews be categorized?}
  \begin{itemize}
     \item  \textit{Motivation.} As aforementioned, IoT product reviews can contain a wide range of information. This information includes, but is not limited to: subjective product evaluations (about hardware, software, or the product as a whole), personal experiences, problems encountered, product descriptions, and users' visions on how the product should be. In this question, we investigate what types of information are in the reviews.
     \end{itemize}
  \item[] \textbf{RQ2: How much information in the studied product reviews is relevant to software engineers?}
	\begin{itemize}
     	\item  \textit{Motivation.} Once the taxonomy is defined in RQ1, the natural next step is to manually examine the reviews and to investigate the distribution of each category, \textit{i.e.}, how often each category occurs. Specifically, we aim to find out how much of this data can help to enable a more responsive software requirements elicitation and software evolution process.
     \end{itemize}
  \item[] \textbf{RQ3: How effective is a machine learning approach in classifying the reviews?}
  \begin{itemize}
     	\item  \textit{Motivation.} Sifting through each review manually to find the right information is a laborious task. Since SVM and CNN are often used for classification tasks, we investigate the extent to which they can be used to classify an unseen review based on the taxonomy previously defined in RQ1 automatically.
        \end{itemize}
\end{itemize}
\subsection{Data Collection}
To conduct our analysis, we sampled 1491 verified purchase reviews from 6 IoT products from Amazon. The data was collected on September 22, 2017 through a web crawling tool we developed. Table \ref{productdetails} describes the characteristics of the 6 IoT products we considered.

\begin{table}[H]
\centering
\caption{Details of the studied products}
\label{productdetails}
\scalebox{0.85}{
\setlength{\tabcolsep}{0.5em} 
\renewcommand{\arraystretch}{1}
\rowcolors{1}{gray!5}{}
\begin{tabular}{llrr}
\toprule
\multicolumn{1}{c}{Product name} & \multicolumn{1}{c}{ASIN} & \multicolumn{1}{c}{Domain} & \multicolumn{1}{c}{Total Reviews} \\ \midrule
Amazon Echo Dot                  & B01DFKC2SO               & Smart Home           & 54230                             \\
Fitbit Blaze                     & B019VM3F2M               & Smart Watch      & 7349                              \\
GoPro Hero4 Silver               & B00NIYJF6U               & Action Camera            & 2327                              \\
PlayStation 4                    & B01LRLJV28               & Gaming Console           & 3135                              \\
Pebble Time                      & B0106IS5XY               & Smart Watch              & 1577                              \\
Amazon Echo Show                 & B01J24C0TI               & Smart Home           & 2565       \\          
\bottomrule
\end{tabular}}
\end{table}

To mitigate some threats to the external validity, we selected products from various domains: smart home, smart watch, gaming console, and action camera. It is important to note that on Amazon, a user can subjectively give a star rating on a scale of 1 to 5. However, as far as we know, there is no clear cut definition to what each star in a five-star rating system means. Furthermore, there is no limit on how many characters a user can write for a review. Therefore, as to eliminate some bias in the dataset, we sampled 50 reviews per each star rating of each product (total of 250 per product) where each review consists of no more than 20 sentences. We used NLTK\footnote{http://www.nltk.org/}'s sentence tokenizer to split a review into a list of sentences. However, at the time of our data collection, 2-star reviews of PlayStation 4 and of GoPro Hero4 Silver contain less than 50 reviews that meet the aforementioned criteria. In this case, we only fetched reviews that meet the criteria from them, 46 and 45 reviews respectively. Table \ref{sentencebreakdown} shows the number of sentences for each star rating of each product. For replicability, our dataset is publicly available here\footnote{https://goo.gl/M6BcB1 }.

\begin{table}[H]
\centering
\caption{Number of sentences per star-rating of each product}
\label{sentencebreakdown}
\scalebox{0.82}{
\setlength{\tabcolsep}{0.4em} 
\begin{tabular}{clcccccr}
\toprule
\multirow{2}{*}{\textbf{\#}} & \multicolumn{1}{l}{\multirow{2}{*}{\textbf{Product}}} & \multicolumn{5}{c}{\textbf{Star rating}}                                                                                                                                                        & \multirow{2}{*}{\textbf{Total}} \\ \cmidrule(lr){3-7}
                             & \multicolumn{1}{l}{}                                  & \multicolumn{1}{l}{\textbf{1-star}} & \multicolumn{1}{l}{\textbf{2-star}} & \multicolumn{1}{l}{\textbf{3-star}} & \multicolumn{1}{l}{\textbf{4-star}} & \multicolumn{1}{l}{\textbf{5-star}} &                                 \\ \midrule
1& Amazon Echo Dot & 155 & 282 & 280 & 385 & 354 & 1456 \\
2& Fitbit Blaze& 234 & 290 & 313 & 361 & 350 & 1548 \\
3& GoPro Hero 4 Silver& 228 & 196 & 165 & 278 & 357 & 1224 \\
4& PlayStation 4 & 134 & 183 & 146 & 99 & 158 & 720 \\
5& Pebble Watch  & 167 & 189 & 189 & 267 & 336 & 1148 \\
6& Amazon Echo Show & 235 & 254 & 149 & 204 & 260 & 1102 \\ \midrule
\multicolumn{7}{r}{Total sentences:} & 7198 \\ \bottomrule
\end{tabular}}
\end{table}
\section{Methodology and Results}
\subsection{RQ1}
\noindent\textbf{Procedure:}

In this section, we discuss the procedure we used to answer \textbf{RQ1: How can IoT product reviews be categorized?} 

We started by searching past literature for similar efforts on analyzing user reviews on IoT products. However, what we found was that most past literature limits its scope by focusing on analyzing user reviews of mobile applications on App Stores, such as Google Play or the Apple Store \cite{khalid2015mobile}\cite{maalej2015bug}, and tweets on Twitter \cite{guzman2016needle}. As a result, user reviews on firmware, operating systems, or embedded software are not explored, unless reviews of a product using them are investigated. In doing so, we cannot assume that the majority of review sentences are software related. Therefore, a taxonomy defined by most past literature would not work.

\textbf{Top-level taxonomy}
We conducted three \textit{content analysis} sessions \cite{morgan1993qualitative} in order to see what information or pattern was contained in the reviews. Upon initial data inspection involving a few review samples, the first author came up with a rudimentary taxonomy which consists of 7 categories: \textit{User Background, Usage Scenario, Software Feature Request, Software Complaint, Software Praise, Hardware Complaint}, and \textit{Hardware Praise}. This resulted in our first draft taxonomy. In the next step, we assigned 52 Master's level CS students, who were taking a graduate level software engineering course at USC, to analyze and categorize IoT product review sentences based on the first draft taxonomy as the course's requirements elicitation activity\footnote{https://goo.gl/wdAtUC}. Note that students could suggest a category if they felt that a review sentence cannot be classified into any categories in the first draft taxonomy. The responses were gathered and new categories suggested by students were noted. In the third step, a team of 4 Master's student, who were not part of the students in the second step, and one PhD student, the first author, validated categories suggested in the second step by merging similar categories together and redefining them. The taxonomy resulting from this step consisted of 8 top-level categories, shown in Table \ref{BigCategory}. Additionally and initially, two of the top-level categories (Hardware and Software) had 5 additional sub-categories: \textit{Praise, Complaint, Inquiry, Request}, and \textit{Other}. However, through examining software categories that are relevant to software engineers from past literature discussed in detail in the next paragraph, we adopted its software taxonomy instead of our original 5 sub-categories. We adapted the sub-categories for Hardware category in the same way. 

\textbf{Software-level taxonomy} In this level, we can assume that all review sentences are software related. Therefore, we could use the existing taxonomy from past literature. We used the \textit{Taxonomy for Software Maintenance and Evolution} purposed by \cite{panichella2015can} as a starting point because they already evaluated the relevance of each category for developers performing software evolution and maintenance tasks. However, we modified the definitions to better reflect the nature of our data. In addition, upon inspecting the data, we found that sentences expressing dissatisfaction with the software of a product often contained information describing why the user was dissatisfied. We believe this is relevant to software developers and requirement engineers in order to better meet user needs or expectations. We added these sentences to the Problem Discovery category. On the other hand, sentences expressing satisfaction conveyed little value for software evolution process \cite{panichella2015can}. In this case, we combined these sentences with the Information Giving category. Table \ref{softwarelevel} shows the 4 sub-categories for Software category used in our study.

\begin{table}[H]
\centering
\resizebox{0.9\textwidth}{!}{%
\begin{threeparttable}
\caption{8 Top-level categories and their respective definitions}
\label{BigCategory}
\setlength{\tabcolsep}{0.3em} 
\renewcommand{\arraystretch}{1.1}
\rowcolors{1}{gray!4}{}
\begin{tabular}[t]{llp{30em}}
\toprule
\textbf{\#} & \textbf{Category}         & \textbf{Definition} \\ \midrule
1      & Hardware        & Sentences mentioning a hardware component, a physical characteristic, or a physical part of the product. \textit{\textbf{Note:}} can be broken down into 4 sub-categories\textbf{*}.             \\
2      & Software         &  Sentences mentioning a product's functionality, a set of capabilities, or GUI. \textit{\textbf{Note:}} can be broken down. See Table \ref{softwarelevel}.          \\
3      & General          &  Sentences describing the product as a whole, including persuading to and dissuading against buying or using the product.           \\
4      & User Background  &  Sentences discussing the background of the reviewer or the reviewer's character.           \\
5      & Other Product    &  Sentences referring to another product (\textit{e.g.,} competitive products, accessories, products that should be used with this product, etc.).          \\
6      & Usage Scenario   &  Sentences mentioning a way to use the product, \textit{i.e.,} a use case scenario.           \\
7      & Customer Service &  Sentences recounting on the reviewer's experience with Amazon services (\textit{e.g.,} package, return, shipment, etc).           \\
8      & Miscellaneous    &   Sentences that are not covered by or do not belong to another category.       \\
\bottomrule
\end{tabular}
\begin{tablenotes}
      \small
      \item \textbf{*}the definition of each sub-category is similar to the definitions provided in Table \ref{softwarelevel}, but with Hardware oriented terms.
    \end{tablenotes}
  \end{threeparttable}}
\end{table}

\begin{table}[H]
\centering
\caption{Categories and their respective definitions for software-related sentences}
\label{softwarelevel}
\rowcolors{1}{gray!4}{}
\setlength{\tabcolsep}{0.4em} 
\renewcommand{\arraystretch}{1.2}
\resizebox{0.9\textwidth}{!}{%
\begin{tabular}{lp{30em}}
\toprule
\textbf{Category}           & \textbf{Definition}                                                                                                           \\\midrule
Information Giving & Sentences expressing satisfaction or informing other users or the sellers about the functionality of the product       \\
Inquiry            & Sentences related to attempts to obtain information or help from other users or the seller about the software \\
Feature Request    & Sentences expressing ideas, suggestions, or needs for improving or enhancing the software or functionality of the product               \\
Problem Discovery  & Sentences expressing dissatisfaction or describing issues or unexpected behaviors with the software of the product   \\ \bottomrule
\end{tabular}}
\end{table}

\noindent\textbf{Results:}

The resulted taxonomy consisting of 8 top-level categories and their respective definitions is shown in Table \ref{BigCategory}. The second level of the Software category is shown in Table \ref{softwarelevel}. This layer is also applicable to the Hardware category with hardware-oriented definitions.

\subsection{RQ2}
\noindent\textbf{Procedure:}

In this section, we discuss the procedure and applied the taxonomy defined in RQ1 to answer \textbf{RQ2: How much information in the studied product reviews is relevant to software engineers?}
\\

\textbf{Manual Classification} To create the ground truth dataset for our experiment, we manually classified each sentence in the dataset. A team consists of 4 Master's level CS students and one PhD students, the first author, accomplished this task. To mitigate some threats to the internal validity, we performed the following procedures:

First, we created a category guide, which consists of the precise details of the category definitions and examples. Second, we conducted a pilot run by randomly selecting 80 reviews from the dataset and having each annotator annotate the selected reviews. We calculated the inter-rater reliability to assess validity and subjectivity of this pilot run using Fleiss's Kappa since we have more than two annotators \cite{fleiss1971measuring}. The inter-rater reliability value was 0.65, which is in the \textit{``Substantial Agreement"} range according to \cite{landis1977measurement}, one level below \textit{``Almost Perfect Agreement"} range. Third, we held several meetings before actual classifying the data to make sure that all annotators had a mutual understanding of the concept regarding the definitions of the categories in the taxonomy and the descriptions of the products. Fourth, we instructed each annotator to annotate at most 100 sentences per day as to avoid errors due to fatigue. Fifth, the first author individually performed a quality check after the dataset was labeled by randomly selecting 50\% of the entire dataset. Any inconsistencies detected, \textit{i.e.,} assigned categories that did not adhere to the definitions in the taxonomy for that sentence, were reviewed and fixed. Sixth, categories of software-level categories were classified by only one annotator. This total process spanned 4 weeks. Table \ref{annotationexamples} shows examples of sentences and their corresponding categories from the manual classification process.
\begin{table}[H]
\centering
\setlength{\tabcolsep}{0.4em} 
\renewcommand{\arraystretch}{1.1}
\caption{Examples of sentences classified by using the taxonomy in RQ1}
\label{annotationexamples}
\rowcolors{1}{gray!5}{}
\resizebox{\textwidth}{!}{%
\begin{tabular}{p{30em}|l|l}
\toprule
\textbf{Review Sentence}                                                                                                                             & \textbf{\begin{tabular}[c]{@{}l@{}}Top-level\\ category\end{tabular}} & \textbf{Second level}                                                        \\ \toprule
``I developed a Class IV allergic reaction to the wrist band."                                                                                & Hardware                                                              & \begin{tabular}[t]{@{}l@{}}Hardware\\ Problem Discovery\end{tabular}          \\
``The companion android app is crap though - very slow to start and not very functional."                          & \begin{tabular}[t]{@{}l@{}}Software\end{tabular}          & \begin{tabular}[t]{@{}l@{}}Software \\Problem Discovery\end{tabular} \\
``Stay away from this product!" & \begin{tabular}[t]{@{}l@{}}General\end{tabular}   &                                                                               \\
``Alexa does not answer general questions as Google Home seems to."                                                                            & \begin{tabular}[t]{@{}l@{}}Software,\\ Other Product\end{tabular}     & \begin{tabular}[t]{@{}l@{}}Software\\ Information Giving\end{tabular}         \\
``I got the camera on time but the LCD screen is not working."                                                                                 & \begin{tabular}[t]{@{}l@{}}Customer Service, \\ Hardware\end{tabular} & \begin{tabular}[t]{@{}l@{}}Hardware\\ Problem Discovery
\end{tabular}\\
``The software needs work: i am constantly having to uninstall reinstall the pebble time app for android."                                                                                 & Software                                                              & \begin{tabular}[t]{@{}l@{}}Software\\ Problem Discovery\end{tabular}          \\
``If I could set my alarm to wake me up M-F @ 7 am that would be a 5 star moment."                                                                                 & Software                                                              & \begin{tabular}[t]{@{}l@{}}Software\\ Feature Request\end{tabular}          \\
``Bought this as a birthday present for my daughter a month ago and gave it to her last night."                                                                                 & User Background                                                              & \begin{tabular}[t]{@{}l@{}}\\ \end{tabular}          \\
\bottomrule   
\end{tabular}}
\end{table}

\noindent\textbf{Results:}

The results of the manual classification process for the top-level categories is shown in Table \ref{Topcat}. Only 26.72\% of the review sentences were found to be software related which means that the majority of review sentences are not directly applicable to software engineers. Table \ref{softwaremanual} shows the manual classification results of only software related sentences by applying the taxonomy in Table \ref{softwarelevel}. Only 8.79\% (2.35\% of all sentences) contain feedback on how to improve or enhance the software, 45.40\% (12.12\% of all sentences) contain information about problems users encountered or what made users dissatisfied with the software, and 1.09\% (0.29\% of all sentences) express an effort to acquire information about the software. This means that only 14.79\% of all sentences are directly applicable to software engineers. These findings answer \textbf{RQ2}.
\begin{table}[H]
\centering
\scalebox{0.9}{
\begin{threeparttable}
\caption{Manual Classification Results for Top-level categories}
\label{Topcat}
\setlength{\tabcolsep}{2.4em} 
\rowcolors{1}{gray!4}{}
\begin{tabular}{@{}lcc@{}}
\toprule
\textbf{Category} & \textbf{Count*} & \textbf{Frequency \%} \\ \midrule
Hardware          & 1870           & 25.98\%               \\
Software          & 1923           & 26.72\%               \\
General           & 2290           & 31.81\%               \\
User Background   & 1711           & 23.77\%               \\
Other Product     & 549            & 7.63\%                \\
Usage Scenario    & 504            & 7.00\%                \\
Customer Service  & 199            & 2.76\%                \\
Miscellaneous     & 291            & 4.04\%                \\ \bottomrule
\end{tabular}
\begin{tablenotes}
      \small
      \item \textbf{*}Each sentence can be classified into one or more top-level categories
    \end{tablenotes}
  \end{threeparttable}}
\end{table}

\begin{table}[H]
\centering
\setlength{\tabcolsep}{1.0em} 
\renewcommand{\arraystretch}{1.05}
\rowcolors{1}{gray!4}{}
\caption{Manual Classification Results for software related sentences}
\label{softwaremanual}
\scalebox{0.9}{
\begin{tabular}{@{}lccc@{}}
\toprule
\textbf{Category}  & \textbf{Count} & \multicolumn{1}{c}{\textbf{\begin{tabular}[c]{@{}c@{}}Frequency \%\\ (software only)\end{tabular}}} & \multicolumn{1}{c}{\textbf{\begin{tabular}[c]{@{}c@{}}Frequency \%\\ (all sentences)\end{tabular}}} \\ \midrule
Information Giving & 860     &   44.72\%    & 11.94\%      
\\
Inquiry            & 21      &    1.09\%   & 0.29\%   
\\
Feature Request    & 169     &   8.79\%    & 2.35\%                \\
Problem Discovery  & 873     &   45.40\%    & 12.12\%               \\\bottomrule
\end{tabular}}
\end{table}

\subsection{RQ3}
\noindent\textbf{Procedure:}

In this section, we discuss the background and procedure to answer \textbf{RQ3: How effective is a machine learning approach in classifying the reviews?} 
In particular, we can break this research question down into three sub-questions: first, what is the performance of each supervised machine learning technique (SVM vs CNN), second, what combination gives us the highest precision and recall, and third, how generalizable is the model.
			
In contrast with multi-class classification where each sentence is assigned to only one label, in our top-level categories, each sentence can be classified into one or more categories in the top-level. Such a problem is referred to as multi-label classification.\\

\textbf{Text preprocessing} We preprocessed the reviews in the following orders: (1) utilizing sentence tokenizer provided by NLTK, a widely used natural language processing toolkit in Python, to breakdown reviews into sentences, (2) lowercasing the resulted sentences, and (3) tokenizing the words on space and punctuation, removing any non-English character. \\

\textbf{Vector Space Model}:

\textit{Term frequency-inverse document frequency scheme (tf-idf)} Tf-idf consists of two components: the term frequency and the inverse document frequency. The former considers the number of occurrences of a term in a document, while the latter takes into consideration the information on the frequency of a term in the entire corpus. In other words, the scheme reflects how important a term is to a document in a corpus \cite{salton1986introduction}. This scheme has been widely used in information retrieval and text mining. 


\textit{Word2Vec (W2V)}. This is based on the distributional hypothesis, which states that words that appear in a similar context tend to share similar meanings \cite{harris1954distributional}. It learns to group similar words together in a vector space in an unsupervised manner. The algorithms to generate vector representations of words are described in detail in \cite{mikolov2013efficient}. It has been widely found that, with enough data and context, word2vec model can improve the performance of a classifier \cite{kim2014convolutional}. To harness the power of this model, we trained our word2vec model on 2.4 million Amazon reviews (over 12 million sentences) based on the dataset made available by \cite{mcauley2013hidden}. Note that we applied the aforementioned steps for text preprocessing before training a word2vec model. We only used reviews from “Electronics” and “Apps for Androids” category.\\

\textbf{Implementation:}

\textit{Support Vector Machines} We used Scikit-learn\footnote{http://scikit-learn.org/stable/modules/svm.html}, a widely used machine learning library for Python. We adopted the method based on combining SVM with tf-idf and SVM with W2V to classify the information presented in the reviews. With SVM + W2V, we selected certain features according to the part-of-speech of words, meaning that we considered only nouns, verbs, adjectives, and adverbs in a given sentence. We then combined vectors of these words and averaged them. In addition, we leveraged binary relevance method as a strategy to transform each label into an independent binary classification problem and train one classifier for each label \cite{luaces2012binary}.

\textit{Convolutional Neural Network} We used TensorFlow\footnote{https://www.tensorflow.org/}, an open-source library for machine intelligence, to implement our CNN model based on Kim's CNN non-static model \cite{kim2014convolutional}. To make our CNN multi-label compatible, we made changes to the architecture as follows: first, we used sigmoid activation function instead of softmax at the output layer; second, we used \textit{sigmoid cross entropy with logits} as our loss function; and third, since the predicted outputs are a set of probabilities, we used a simple rounding function to convert probabilities into one's and zero's in order to evaluate the accuracy against the ground truth.
\\

\textbf{Metrics} Evaluating the performance of multi-label classification classifier is more complicated than that of multi-class classification. In multi-label classification, predictions can be fully correct (exact match), partially correct (partial match), or fully incorrect (none match) \cite{sorower2010literature}. Therefore, several evaluation metrics should be reported. In this paper, we reported the performance of our classifier on 2 label-based metrics: Macro-average Precision and Macro-averaged Recall. However, the complete results for Example-based metrics (Hamming Loss, Jaccard Similarity, etc) can also be found in our project repository.


\textit{Label-based} measures evaluate each label separately and then averages over all labels \cite{sorower2010literature}.
\begin{equation}
Precision_{MacroAvg} = \frac{1}{n}\sum_{k=1}^{n}\frac{TP_{k}}{TP_{k}+FP_{k}}
\end{equation}
\begin{equation}
Recall_{MacroAvg} = \frac{1}{n}\sum_{k=1}^{n}\frac{TP_{k}}{TP_{k}+FN_{k}}
\end{equation}
where \textit{n} is the number of labels, \textit{TP\textsubscript{k}} is the total number of instances that are correctly identified by the approach for the label k, \textit{FP\textsubscript{k}} is the total number of instances incorrectly identified by the approach for the label k, and \textit{FN\textsubscript{k}} is the total number of true instances that are not identified by the approach for the label k.

\noindent\textbf{Results:}

We decided to group category (3) User Background, (4) Other Product, (5) Usage Scenario, (6) Customer Service, and (7) Miscellaneous together because we believe that correctly identifying the information about the product (Hardware, Software, and General) itself matters the most. Another reason is that some of these categories occur much less frequently. A model that only returns predictions for categories that occur more frequently and never returns categories that occur less frequently will have high accuracy. However, this does not indicate that the model has a good predictive power as the model does not return any other label, except the most frequent one. This type of model is often used as a simple baseline model. For the sub-categories, we also decided to group category Inquiry with Problem Discovery because Inquiry occur much less frequently than any other category (Table \ref{softwarelevelresults}). However, we decided not to group Feature Request category with any other category since this category is crucial for software engineers performing software evolution and maintenance tasks. 

Furthermore, to prevent over-fitting and better test the generalizability of the model, we adopted two cross-validation measures. The results reported for the above metrics are based on these cross-validation measures.

\textbf{10-fold cross validation} We used the standard 10-fold cross validation to split the dataset in 10 folds and used 9 folds for training and 1 fold for evaluating. This process is repeated 10 times, rotating the training and testing folds.

\textbf{6-fold product cross-validation} Reviews of each product may contain words and jargon only applicable to the product. That means that term-features that work for one product may not work for others. We conducted a cross-product validation, a method previously proposed by \cite{bacchelli2012content}. In particular, we divided the dataset into 6 folds where each fold represents reviews of each product. We trained the classifier on 5 products reviews and tried to predict the classification of the reviews in the remaining product. The process is repeated 6 times, rotating the training and testing folds.

\begin{table}[H]
\centering
\caption{Classification performance on the top level categories}
\label{toplevelresults}
\setlength{\tabcolsep}{0.12em} 
\resizebox{\textwidth}{!}{%
\begin{threeparttable}
\begin{tabular}{|c|
>{\columncolor[HTML]{E7E7E7}}c |
>{\columncolor[HTML]{FFFFFF}}c |
>{\columncolor[HTML]{E7E7E7}}c |
>{\columncolor[HTML]{FFFFFF}}c |
>{\columncolor[HTML]{E7E7E7}}c |
>{\columncolor[HTML]{FFFFFF}}c |
>{\columncolor[HTML]{E7E7E7}}c |
>{\columncolor[HTML]{FFFFFF}}c |
>{\columncolor[HTML]{E7E7E7}}c |c|
>{\columncolor[HTML]{E7E7E7}}c |c|
>{\columncolor[HTML]{E7E7E7}}c |c|
>{\columncolor[HTML]{E7E7E7}}c |c|
>{\columncolor[HTML]{E7E7E7}}c |
>{\columncolor[HTML]{FFFFFF}}c |
>{\columncolor[HTML]{E7E7E7}}c |
>{\columncolor[HTML]{FFFFFF}}c |
>{\columncolor[HTML]{E7E7E7}}c |
>{\columncolor[HTML]{FFFFFF}}c |
>{\columncolor[HTML]{E7E7E7}}c |
>{\columncolor[HTML]{FFFFFF}}c |}
\hline
\multicolumn{1}{|l|}{} & \multicolumn{24}{c|}{\cellcolor[HTML]{FFFFFF}\textbf{6-fold Product Cross Validation}} \\ \cline{2-25} 
\multicolumn{1}{|l|}{} & \multicolumn{8}{c|}{\cellcolor[HTML]{E7E7E7}SVM + tf-idf\textbf{*}} & \multicolumn{8}{c|}{\cellcolor[HTML]{FFFFFF}SVM + w2v$\ddagger$} & \multicolumn{8}{c|}{\cellcolor[HTML]{E7E7E7}CNN + w2v$\ddagger$} \\ \cline{2-25} 
\multicolumn{1}{|l|}{} & \multicolumn{2}{c|}{\cellcolor[HTML]{E7E7E7}HW} & \multicolumn{2}{c|}{\cellcolor[HTML]{E7E7E7}SW} & \multicolumn{2}{c|}{\cellcolor[HTML]{E7E7E7}GN} & \multicolumn{2}{c|}{\cellcolor[HTML]{E7E7E7}OT} & \multicolumn{2}{c|}{\cellcolor[HTML]{FFFFFF}HW} & \multicolumn{2}{c|}{\cellcolor[HTML]{FFFFFF}SW} & \multicolumn{2}{c|}{\cellcolor[HTML]{FFFFFF}GN} & \multicolumn{2}{c|}{\cellcolor[HTML]{FFFFFF}OT} & \multicolumn{2}{c|}{\cellcolor[HTML]{E7E7E7}HW} & \multicolumn{2}{c|}{\cellcolor[HTML]{E7E7E7}SW} & \multicolumn{2}{c|}{\cellcolor[HTML]{E7E7E7}GN} & \multicolumn{2}{c|}{\cellcolor[HTML]{E7E7E7}OT} \\ \cline{2-25} 
\multicolumn{1}{|l|}{\multirow{-4}{*}{Product\#}} & P & \cellcolor[HTML]{E7E7E7}R & P & \cellcolor[HTML]{E7E7E7}R & P & \cellcolor[HTML]{E7E7E7}R & P & \cellcolor[HTML]{E7E7E7}R & P & \cellcolor[HTML]{FFFFFF}R & P & \cellcolor[HTML]{FFFFFF}R & P & \cellcolor[HTML]{FFFFFF}R & P & \cellcolor[HTML]{FFFFFF}R & P & \cellcolor[HTML]{E7E7E7}R & P & \cellcolor[HTML]{E7E7E7}R & P & \cellcolor[HTML]{E7E7E7}R & P & \cellcolor[HTML]{E7E7E7}R \\ \hline
1 & 0.61 & 0.53 & 0.75 & 0.52 & 0.65 & 0.56 & 0.65 & 0.63 & 0.63 & \cellcolor[HTML]{FFFFFF}0.60 & 0.75 & \cellcolor[HTML]{FFFFFF}0.58 & 0.70 & \cellcolor[HTML]{FFFFFF}0.56 & 0.73 & \cellcolor[HTML]{FFFFFF}0.64 & 0.70 & 0.43 & 0.75 & 0.56 & 0.76 & 0.51 & 0.68 & 0.66 \\ \hline
2 & 0.75 & 0.59 & 0.71 & 0.55 & 0.62 & 0.60 & 0.66 & 0.51 & 0.76 & \cellcolor[HTML]{FFFFFF}0.61 & 0.74 & \cellcolor[HTML]{FFFFFF}0.61 & 0.72 & \cellcolor[HTML]{FFFFFF}0.62 & 0.73 & \cellcolor[HTML]{FFFFFF}0.55 & 0.79 & 0.56 & 0.75 & 0.62 & 0.70 & 0.53 & 0.70 & 0.54 \\ \hline
3 & 0.75 & 0.59 & 0.53 & 0.46 & 0.68 & 0.47 & 0.64 & 0.65 & 0.76 & \cellcolor[HTML]{FFFFFF}0.65 & 0.53 & \cellcolor[HTML]{FFFFFF}0.59 & 0.75 & \cellcolor[HTML]{FFFFFF}0.39 & 0.67 & \cellcolor[HTML]{FFFFFF}0.68 & 0.78 & 0.62 & 0.59 & 0.53 & 0.77 & 0.46 & 0.78 & 0.52 \\ \hline
4 & 0.70 & 0.35 & 0.46 & 0.43 & 0.67 & 0.53 & 0.68 & 0.56 & 0.80 & \cellcolor[HTML]{FFFFFF}0.44 & 0.51 & \cellcolor[HTML]{FFFFFF}0.43 & 0.72 & \cellcolor[HTML]{FFFFFF}0.58 & 0.70 & \cellcolor[HTML]{FFFFFF}0.73 & 0.85 & 0.34 & 0.53 & 0.45 & 0.70 & 0.54 & 0.72 & 0.67 \\ \hline
5 & 0.83 & 0.66 & 0.71 & 0.65 & 0.72 & 0.58 & 0.57 & 0.53 & 0.83 & \cellcolor[HTML]{FFFFFF}0.65 & 0.74 & \cellcolor[HTML]{FFFFFF}0.60 & 0.75 & \cellcolor[HTML]{FFFFFF}0.61 & 0.64 & \cellcolor[HTML]{FFFFFF}0.58 & 0.85 & 0.64 & 0.70 & 0.71 & 0.75 & 0.61 & 0.64 & 0.52 \\ \hline
6 & 0.71 & 0.60 & 0.75 & 0.56 & 0.66 & 0.48 & 0.59 & 0.58 & 0.77 & \cellcolor[HTML]{FFFFFF}0.57 & 0.78 & \cellcolor[HTML]{FFFFFF}0.66 & 0.77 & \cellcolor[HTML]{FFFFFF}0.57 & 0.66 & \cellcolor[HTML]{FFFFFF}0.65 & 0.74 & 0.62 & 0.82 & 0.61 & 0.72 & 0.62 & 0.70 & 0.57 \\ \hline
\cellcolor[HTML]{E7E7E7}\textbf{P(MA)} & \multicolumn{8}{c|}{\cellcolor[HTML]{E7E7E7}0.67} & \multicolumn{8}{c|}{\cellcolor[HTML]{E7E7E7}0.71} & \multicolumn{8}{c|}{\cellcolor[HTML]{E7E7E7}0.73} \\ \hline
\textbf{R(MA)} & \multicolumn{8}{c|}{\cellcolor[HTML]{FFFFFF}0.55} & \multicolumn{8}{c|}{\cellcolor[HTML]{FFFFFF}0.59} & \multicolumn{8}{c|}{\cellcolor[HTML]{FFFFFF}0.56} \\ \hline
\multicolumn{25}{|c|}{\cellcolor[HTML]{FFFFFF}\textbf{10 fold Cross Validation}} \\ \hline
\multicolumn{1}{|l|}{} & 0.77 & 0.63 & 0.72 & 0.60 & 0.69 & 0.59 & 0.65 & 0.60 & 0.77 & 0.65 & 0.74 & 0.64 & 0.75 & 0.59 & 0.71 & 0.64 & 0.80 & 0.62 & 0.77 & 0.61 & 0.73 & 0.61 & 0.72 & 0.60 \\ \hline
\cellcolor[HTML]{E7E7E7}\textbf{P(MA)} & \multicolumn{8}{c|}{\cellcolor[HTML]{E7E7E7}0.71} & \multicolumn{8}{c|}{\cellcolor[HTML]{E7E7E7}0.74} & \multicolumn{8}{c|}{\cellcolor[HTML]{E7E7E7}0.76} \\ \hline
\textbf{R(MA)} & \multicolumn{8}{c|}{\cellcolor[HTML]{FFFFFF}0.60} & \multicolumn{8}{c|}{\cellcolor[HTML]{FFFFFF}0.63} & \multicolumn{8}{c|}{\cellcolor[HTML]{FFFFFF}0.61} \\ \hline
\end{tabular}
\begin{tablenotes}
      \small
      \item HW = Hardware; SW = Software; GN = General; OT = Other; P = Precision; R = Recall; P(MA) = Macro Average Precision; R(MA) = Macro Average Recall; \textbf{*} included stop words; $\ddagger$ trained on the combination of 2.4 millions Amazon reviews with our dataset
    \end{tablenotes}
  \end{threeparttable}}
\end{table}

\begin{table}[H]
\centering
\caption{Classification performance on software related categories}
\label{softwarelevelresults}
\setlength{\tabcolsep}{0.4em} 
\resizebox{\textwidth}{!}{%
\begin{threeparttable}
\begin{tabular}{|c|c|c|c|c|c|c|c|c|c|c|c|c|c|c|c|c|c|c|}
\hline
 & \multicolumn{18}{c|}{\textbf{6 fold Product Cross Validation}} \\ \cline{2-19} 
 & \multicolumn{6}{c|}{\cellcolor[HTML]{E7E7E7}SVM + tf-idf\textbf{*}} & \multicolumn{6}{c|}{SVM + w2v$\ddagger$} & \multicolumn{6}{c|}{\cellcolor[HTML]{E7E7E7}CNN + w2v$\ddagger$} \\ \cline{2-19} 
 & \multicolumn{2}{c|}{\cellcolor[HTML]{E7E7E7}FR} & \multicolumn{2}{c|}{\cellcolor[HTML]{E7E7E7}IG} & \multicolumn{2}{c|}{\cellcolor[HTML]{E7E7E7}PD} & \multicolumn{2}{c|}{FR} & \multicolumn{2}{c|}{IG} & \multicolumn{2}{c|}{PD} & \multicolumn{2}{c|}{\cellcolor[HTML]{E7E7E7}FR} & \multicolumn{2}{c|}{\cellcolor[HTML]{E7E7E7}IG} & \multicolumn{2}{c|}{\cellcolor[HTML]{E7E7E7}PD} \\ \cline{2-19} 
\multirow{-4}{*}{Product\#} & \cellcolor[HTML]{E7E7E7}P & \cellcolor[HTML]{E7E7E7}R & \cellcolor[HTML]{E7E7E7}P & \cellcolor[HTML]{E7E7E7}R & \cellcolor[HTML]{E7E7E7}P & \cellcolor[HTML]{E7E7E7}R & P & R & P & R & P & R & \cellcolor[HTML]{E7E7E7}P & \cellcolor[HTML]{E7E7E7}R & \cellcolor[HTML]{E7E7E7}P & \cellcolor[HTML]{E7E7E7}R & \cellcolor[HTML]{E7E7E7}P & \cellcolor[HTML]{E7E7E7}R \\ \hline
1 & \cellcolor[HTML]{E7E7E7}0.73 & \cellcolor[HTML]{FFFFFF}0.41 & \cellcolor[HTML]{E7E7E7}0.71 & \cellcolor[HTML]{FFFFFF}0.58 & \cellcolor[HTML]{E7E7E7}0.64 & \cellcolor[HTML]{FFFFFF}0.81 & \cellcolor[HTML]{E7E7E7}0.74 & 0.49 & \cellcolor[HTML]{E7E7E7}0.77 & 0.60 & \cellcolor[HTML]{E7E7E7}0.68 & 0.87 & \cellcolor[HTML]{E7E7E7}0.87 & \cellcolor[HTML]{FFFFFF}0.22 & \cellcolor[HTML]{E7E7E7}0.75 & \cellcolor[HTML]{FFFFFF}0.53 & \cellcolor[HTML]{E7E7E7}0.62 & \cellcolor[HTML]{FFFFFF}0.90 \\ \hline
2 & \cellcolor[HTML]{E7E7E7}0.55 & \cellcolor[HTML]{FFFFFF}0.35 & \cellcolor[HTML]{E7E7E7}0.73 & \cellcolor[HTML]{FFFFFF}0.75 & \cellcolor[HTML]{E7E7E7}0.72 & \cellcolor[HTML]{FFFFFF}0.72 & \cellcolor[HTML]{E7E7E7}0.58 & 0.41 & \cellcolor[HTML]{E7E7E7}0.78 & 0.87 & \cellcolor[HTML]{E7E7E7}0.83 & 0.74 & \cellcolor[HTML]{E7E7E7}1.00 & \cellcolor[HTML]{FFFFFF}0.24 & \cellcolor[HTML]{E7E7E7}0.73 & \cellcolor[HTML]{FFFFFF}0.88 & \cellcolor[HTML]{E7E7E7}0.83 & \cellcolor[HTML]{FFFFFF}0.70 \\ \hline
3 & \cellcolor[HTML]{E7E7E7}0.62 & \cellcolor[HTML]{FFFFFF}0.56 & \cellcolor[HTML]{E7E7E7}0.78 & \cellcolor[HTML]{FFFFFF}0.65 & \cellcolor[HTML]{E7E7E7}0.72 & \cellcolor[HTML]{FFFFFF}0.83 & \cellcolor[HTML]{E7E7E7}0.75 & 0.33 & \cellcolor[HTML]{E7E7E7}0.84 & 0.75 & \cellcolor[HTML]{E7E7E7}0.78 & 0.90 & \cellcolor[HTML]{E7E7E7}0.00 & \cellcolor[HTML]{FFFFFF}0.00 & \cellcolor[HTML]{E7E7E7}0.79 & \cellcolor[HTML]{FFFFFF}0.62 & \cellcolor[HTML]{E7E7E7}0.68 & \cellcolor[HTML]{FFFFFF}0.88 \\ \hline
4 & \cellcolor[HTML]{E7E7E7}0.33 & \cellcolor[HTML]{FFFFFF}0.33 & \cellcolor[HTML]{E7E7E7}0.59 & \cellcolor[HTML]{FFFFFF}0.61 & \cellcolor[HTML]{E7E7E7}0.79 & \cellcolor[HTML]{FFFFFF}0.78 & \cellcolor[HTML]{E7E7E7}0.33 & 0.33 & \cellcolor[HTML]{E7E7E7}0.79 & 0.67 & \cellcolor[HTML]{E7E7E7}0.81 & 0.88 & \cellcolor[HTML]{E7E7E7}0.00 & \cellcolor[HTML]{FFFFFF}0.00 & \cellcolor[HTML]{E7E7E7}0.82 & \cellcolor[HTML]{FFFFFF}0.70 & \cellcolor[HTML]{E7E7E7}0.82 & \cellcolor[HTML]{FFFFFF}0.91 \\ \hline
5 & \cellcolor[HTML]{E7E7E7}0.55 & \cellcolor[HTML]{FFFFFF}0.60 & \cellcolor[HTML]{E7E7E7}0.82 & \cellcolor[HTML]{FFFFFF}0.65 & \cellcolor[HTML]{E7E7E7}0.58 & \cellcolor[HTML]{FFFFFF}0.78 & \cellcolor[HTML]{E7E7E7}0.40 & 0.40 & \cellcolor[HTML]{E7E7E7}0.87 & 0.64 & \cellcolor[HTML]{E7E7E7}0.57 & 0.83 & \cellcolor[HTML]{E7E7E7}1.00 & \cellcolor[HTML]{FFFFFF}0.20 & \cellcolor[HTML]{E7E7E7}0.88 & \cellcolor[HTML]{FFFFFF}0.65 & \cellcolor[HTML]{E7E7E7}0.56 & \cellcolor[HTML]{FFFFFF}0.86 \\ \hline
6 & \cellcolor[HTML]{E7E7E7}0.81 & \cellcolor[HTML]{FFFFFF}0.41 & \cellcolor[HTML]{E7E7E7}0.59 & \cellcolor[HTML]{FFFFFF}0.79 & \cellcolor[HTML]{E7E7E7}0.72 & \cellcolor[HTML]{FFFFFF}0.64 & \cellcolor[HTML]{E7E7E7}0.84 & 0.38 & \cellcolor[HTML]{E7E7E7}0.61 & 0.85 & \cellcolor[HTML]{E7E7E7}0.80 & 0.71 & \cellcolor[HTML]{E7E7E7}0.90 & \cellcolor[HTML]{FFFFFF}0.13 & \cellcolor[HTML]{E7E7E7}0.50 & \cellcolor[HTML]{FFFFFF}0.84 & \cellcolor[HTML]{E7E7E7}0.74 & \cellcolor[HTML]{FFFFFF}0.58 \\ \hline
\rowcolor[HTML]{E7E7E7} 
\textbf{P(MA)} & \multicolumn{6}{c|}{\cellcolor[HTML]{E7E7E7}0.67} & \multicolumn{6}{c|}{\cellcolor[HTML]{E7E7E7}0.71} & \multicolumn{6}{c|}{\cellcolor[HTML]{E7E7E7}0.70} \\ \hline
\textbf{R(MA)} & \multicolumn{6}{c|}{\cellcolor[HTML]{FFFFFF}0.63} & \multicolumn{6}{c|}{\cellcolor[HTML]{FFFFFF}0.65} & \multicolumn{6}{c|}{\cellcolor[HTML]{FFFFFF}0.55} \\ \hline
\multicolumn{19}{|c|}{\textbf{10-fold Cross Validation}} \\ \hline
\multicolumn{1}{|l|}{} & \multicolumn{1}{l|}{\cellcolor[HTML]{E7E7E7}0.76} & \multicolumn{1}{l|}{\cellcolor[HTML]{FFFFFF}0.47} & \multicolumn{1}{l|}{\cellcolor[HTML]{E7E7E7}0.74} & \multicolumn{1}{l|}{\cellcolor[HTML]{FFFFFF}0.76} & \multicolumn{1}{l|}{\cellcolor[HTML]{E7E7E7}0.75} & \multicolumn{1}{l|}{\cellcolor[HTML]{FFFFFF}0.78} & \multicolumn{1}{l|}{\cellcolor[HTML]{E7E7E7}0.72} & \multicolumn{1}{l|}{0.47} & \multicolumn{1}{l|}{\cellcolor[HTML]{E7E7E7}0.75} & \multicolumn{1}{l|}{0.76} & \multicolumn{1}{l|}{\cellcolor[HTML]{E7E7E7}0.75} & \multicolumn{1}{l|}{0.80} & \multicolumn{1}{l|}{\cellcolor[HTML]{E7E7E7}0.79} & \multicolumn{1}{l|}{\cellcolor[HTML]{FFFFFF}0.26} & \multicolumn{1}{l|}{\cellcolor[HTML]{E7E7E7}0.71} & \multicolumn{1}{l|}{\cellcolor[HTML]{FFFFFF}0.80} & \multicolumn{1}{l|}{\cellcolor[HTML]{E7E7E7}0.76} & \multicolumn{1}{l|}{\cellcolor[HTML]{FFFFFF}0.75} \\ \hline
\rowcolor[HTML]{E7E7E7} 
\textbf{P(MA)} & \multicolumn{6}{c|}{\cellcolor[HTML]{E7E7E7}0.75} & \multicolumn{6}{c|}{\cellcolor[HTML]{E7E7E7}0.74} & \multicolumn{6}{c|}{\cellcolor[HTML]{E7E7E7}0.75} \\ \hline
\rowcolor[HTML]{FFFFFF} 
\textbf{R(MA)} & \multicolumn{6}{c|}{\cellcolor[HTML]{FFFFFF}0.67} & \multicolumn{6}{c|}{\cellcolor[HTML]{FFFFFF}0.68} & \multicolumn{6}{c|}{\cellcolor[HTML]{FFFFFF}0.60} \\ \hline
\end{tabular}
\begin{tablenotes}
      \small
      \item FR = Feature Request; IG = Information Giving; PD = Problem Discovery; P = Precision; R = Recall; P(MA) = Macro Average Precision; R(MA) = Macro Average Recall; \textbf{*} included stop words; $\ddagger$ trained on the combination of 2.4 millions Amazon reviews with our dataset
    \end{tablenotes}
  \end{threeparttable}}
\end{table}

The performance results for classifying top-level categories in Table \ref{toplevelresults} show that incorporating Word2Vec improved precision and recall of SVM in both 6-fold and 10-fold. However, the performance of CNN + w2v and SVM + w2v are comparable. The 6-fold cross validation results show that the SVM + w2v and CNN + w2v generalized well, {\textit{i.e.,} their performances did not drop significantly as SVM + tf-idf}. Table \ref{softwarelevelresults} shows the performance of different methods in classifying software sentences into different software categories. Since only a small quantity of Feature Request sentences are in the dataset, the performance of CNN dropped significantly. In fact, for product 3 (GoPro Hero4 Silver) and 4 (PlayStation 4), CNN did not classify any sentence in the Feature Request category. SVM + w2v's performance, on the other hand, did not drop as significantly as CNN + w2v from 10-fold to 6-fold product cross validation. This implies that the combination generalized well across different products. Surprisingly, SVM + tf-idf outperformed CNN + w2v on classifying software related sentence (R(MA) for 10-fold). However, with more software related sentences, we believe that CNN would perform as well as or slightly better than SVM. These findings answer \textbf{RQ3}.

\section{Threats to Validity}
The study presented in this paper has several factors that may affect the validity of the results. In the following, we discuss possible threats to the validity in our study.

\textbf{Taxonomy} This threat concerns the validity of our taxonomy. To mitigate this threat, we conducted both internal content analysis, involving a team of 5, and external content analysis, involving 52 master's level CS students, to analyze what information or pattern contained in the reviews (see Section 3.1). In addition, we also adapted taxonomy proposed by past literature in our work. However, other studies that use taxonomy with a different set of categories and definitions might lead to different results.

\textbf{Subjectivity in manual classification} This threat stems from the fact that our ground truth dataset is based on human judgment. However, it is not uncommon to involve humans to manually classify data in a text classification problem. To mitigate such a threat, we employ multiple measures as aforementioned in Section 3.2. Nevertheless, we cannot claim that our dataset is error-free as some bias may remain.

\textbf{External Validity} This threat concern how generalizable our results are. To mitigate such a threat, we selected products from different application domains: smart home, smart watch, action camera, and gaming console. Furthermore, to eliminate some bias in our dataset, we selected at most 50 reviews per star rating for each product, and each review contains less than 20 sentences. We also performed a project 6-fold cross validation (see Section 3.3) to test the generalizability of our machine learning models. Additionally, since all IoT products have the capability to transfer data or connect to its ecosystem or the Internet, this shared similarity should allow our results to generalize. Nevertheless, we encourage further research to investigate whether our results hold in other IoT domains.

\section{Related Work}
Analyzing user reviews for useful information has gained a lot of attention from researchers in recent years. We highlighted past literature that share similarities with our work.

Pagano and Maalej \cite{pagano2013user}, Khalid et al. \cite{khalid2015mobile}, Hoon et al. \cite{hoon2013analysis}, and Harman et al. \cite{harman2012app} conducted exploratory studies by investigating multiple aspects of user reviews from mobile application distribution platforms such as the Google Store, the Apple Store, or the BlackBerry Store. They found that user reviews contain information that is useful to the developers and the companies. Motivated by their findings, we investigated user reviews on IoT products to see whether enough software related information exists and how much of those information is directly applicable to software engineers.

Guzman et al. \cite{guzman2016needle} conducted an exploratory study by analyzing the content in Twitter's tweets to find useful information for software engineers. They manually classified 1000 tweets and found that tweets contain relevant information for different stakeholders. They also investigated the automation potential by using several supervised machine learning techniques. Our work is similar to their work, however, our studies' subjects are not software applications.
 
Maalej and Nabil \cite{maalej2015bug} compared multiple methods that can help with classifying user reviews on App stores automatically. Similarly, Panichella et al. \cite{panichella2015can} applied several machine learning techniques to classify information in user reviews from App stores. They applied Natural Language Processing (NLP), sentiment analysis, and text analysis to help with classification tasks. They found that combining NLP with sentiment analysis improves both precision and recall significantly. In contrast, we included a neural network approach (CNN) and investigated if vector space models such as Word2Vec and TF-IDF improve the performance of the classifiers. 

\section{Conclusions and Future Work}
In this paper, we conducted an exploratory study by analyzing 1491 verified purchase reviews (7198 review sentences) of 6 IoT products obtained from Amazon. Our results demonstrate that only 26.72\% of all sentences are software related, based on our taxonomy defined through external and internal content analysis sessions. We investigated how much information in those software related sentences is useful for software engineers performing software evolution and maintenance tasks. The results show that only 55.28\% of software related sentences (14.79\% of all sentences) are directly applicable to software engineers. Given that only a small quantity of sentences can help to accelerate software requirements elicitation or evolution process, we studied the extent to which two supervised machine learning techniques (SVM and CNN) can be used to differentiate information contained in the reviews automatically. The results show that utilizing Word2Vec improved the performance of SVM. In addition, Word2Vec also helped the model to generalize better, when classifying unseen reviews of a different product, than tf-idf.

This work can be extended to several directions. For instance, we plan to incorporate more product from several other IoT domains, to include software quality taxonomy to investigate different aspects of the product's software with regards to its software quality, to include feedback from the manufacturers on how useful the findings are, to incorporate sentiment analysis and other types of preprocessing approaches from past literature to further improve the performance of the classifier, to officially compare the performance of our classifier with that of past literature for software-level sentences, and finally to investigate if we can capture and construct a formal requirement (\textit{e.g.,} a user story) from these review sentences automatically.
\section*{Acknowledgement}
This material is based upon work supported in part by the U.S. Department of Defense through the Systems Engineering Research Center (SERC) under Contract H98230-08-D-0171. SERC is a federally funded University Affiliated Research Center managed by Stevens Institute of Technology. It is also supported by the National Science Foundation grant CMMI-1408909, Developing a Constructive Logic-Based Theory of Value-Based Systems Engineering.
\bibliographystyle{splncs03} 

\end{document}